\documentclass[
aps,12pt,final,notitlepage,oneside,onecolumn,nobibnotes,nofootinbib,
noshowpacs,centertags]{revtex4}
\usepackage{graphicx}
\usepackage{dcolumn}
\usepackage{bm}
\usepackage{amsmath,amssymb}
\renewcommand{\vec}[1]{\ensuremath{\bm{#1}}}

\newcommand{\eq}[1]{\begin{equation}#1\end{equation}}

\newcommand{\bra}[1]{\ensuremath{\langle{#1}|\,}}
\newcommand{\ket}[1]{\ensuremath{\,|{#1}\rangle}}

\newcommand{\matrixe}[3]{\ensuremath{\langle{#1}|\,{#2}\,|{#3}\rangle}}

\newcommand{\op}[1]{\ensuremath{\mathrm{#1}}}
\newcommand{\adj}[1]{\ensuremath{{{#1}}^{\dag}}}
\newcommand{\corr}[1]{\ensuremath{\hat{#1}}}

\newcommand{\dd}{\ensuremath{\mathrm{d}}}

\renewcommand{\vec}[1]{\ensuremath{\bm{#1}}}

\newcommand{\gO}{\ensuremath{\op{g}}}

\newcommand{\qO}{\ensuremath{\op{q}}}
\newcommand{\rO}{\ensuremath{\op{r}}}

\newcommand{\CO}{\ensuremath{\op{C}}}
\newcommand{\CCO}{\ensuremath{\adj{\op{C}}}}

\newcommand{\HO}{\ensuremath{\op{H}}}

\newcommand{\OO}{\ensuremath{\op{O}}}

\newcommand{\TO}{\ensuremath{\op{T}}}
\newcommand{\VO}{\ensuremath{\op{V}}}

\newcommand{\rV}{\ensuremath{\vec{r}}}

\newcommand{\qOV}{\ensuremath{\vec{\op{q}}}}
\newcommand{\rOV}{\ensuremath{\vec{\op{r}}}}

\newcommand{\sigmaOV}{\ensuremath{\vec{\op{\sigma}}}}

\newcommand{\Rm}{\ensuremath{R_-}}

\newcommand{\Rp}{\ensuremath{R_+}}

\newcommand{\UCOM}{\ensuremath{\textrm{UCOM}}}
\newcommand{\cm}{\ensuremath{\textrm{cm}}}

\begin{document}
\linespread{1.6}

\title{Collective excitations in the Unitary Correlation Operator Method and
relativistic QRPA studies of exotic nuclei}
\author{N. Paar}
\author{P. Papakonstantinou}
\author{H. Hergert}
\author{R.Roth}
\affiliation{ Institut f\" ur Kernphysik, Technische Universit\" at Darmstadt, Schlossgartenstrasse 9,
D-64289 Darmstadt, Germany}
\date{\today}

\begin{abstract}
The collective excitation phenomena in atomic nuclei are studied in two
different formulations of the Random Phase Approximation (RPA):
(i) RPA based on correlated realistic nucleon-nucleon interactions constructed
within the Unitary Correlation Operator Method (UCOM), and (ii)  relativistic RPA (RRPA)
derived from effective Lagrangians with density-dependent meson-exchange
interactions. The former includes the dominant interaction-induced short-range
central and tensor correlations by means of an unitary transformation. It
is shown that UCOM-RPA correlations induced by collective nuclear vibrations 
recover a part of the residual long-range correlations that are not explicitly
included in the UCOM Hartree-Fock ground state. Both RPA models
are employed in studies of  the isoscalar monopole resonance (ISGMR) in
closed-shell nuclei across the nuclide chart, with an emphasis on the
sensitivity of  its properties  on the constraints for the range of the
UCOM correlation functions.  Within the Relativistic Quasiparticle RPA
(RQRPA) based on Relativistic Hartree-Bogoliubov model, the occurrence
of pronounced low-lying dipole excitations is predicted in nuclei towards the
proton drip-line. From the analysis of the transition densities and the structure
of the RQRPA amplitudes, it is shown that these states correspond to the proton
pygmy dipole resonance.    
\end{abstract}

\maketitle

\bigskip \bigskip

\section{\label{secI}Introduction}

Among various theoretical approaches to nuclear structure, two pathways
have been extensively pursued over the past decades: (i) models based on
effective nuclear interactions constrained by the properties of nuclear
matter and bulk properties of finite nuclei (e.g. Skyrme \cite{Vau.72},
Gogny \cite{Dec.80}, and relativistic models based on exchange of
effective mesons \cite{Vre.05}),  and (ii) models which start from a
realistic nucleon-nucleon (NN) interaction. Recently, several modern
realistic NN interactions have been constructed,  e.g., the Argonne V18
\cite{Wir.95}, the CD-Bonn \cite{Mac.01}, and chiral potentials
\cite{Ent.02}, which reproduce the experimental NN phase-shifts with high
accuracy. Within ab initio  Green's function Monte Carlo \cite{Pie.04} and
no-core shell model calculations \cite{Nav.00} of ground state properties
and low-lying excitation spectra of light nuclei it was shown that 
realistic NN interactions, supplemented by a three-nucleon force, allow for
a quantitative description of experimental data \cite{Pie.01,Epe.02}.

Realistic NN interactions cannot be directly employed in a standard
Hartree-Fock (HF) scheme due to the importance of interaction-induced
correlations in the many-body state beyond the simple HF Slater
determinant. Therefore, an effective, phase-shift equivalent
interaction has to be derived by explicitly accounting for the dominant
correlations. One way to tackle this issue is the Unitary Correlation
Operator Method (UCOM) which describes the short-range central and tensor
correlations by means of an unitary transformation
\cite{Fel.98,Nef.03,Rot.04,Rot.05}. The unitary transformation of the
Hamiltonian including a realistic NN potential results in a correlated
effective interaction well suited for the application with simple
uncorrelated many-body states. An alternative method to derive phase-shift
equivalent, low-momentum effective interactions is the $V_{\text{low}-k}$
renormalization group approach
\cite{Bog.03}.

Studies of collective excitation phenomena in atomic nuclei provide 
valuable insight into many properties of the underlying effective
interactions employed in solving the nuclear many-body problem. In order
to describe  small-amplitude collective excitations within the UCOM
framework, one can employ the random-phase approximation (RPA)
\cite{Row.70}, based on the HF single-nucleon basis.  Since three-body
interactions presently are not included, the results provide information
on their importance for the understanding of collective nuclear
excitations. The UCOM-RPA model can also be employed to evaluate the
contributions of RPA correlations to the ground state energy which go
beyond the mean-field picture \cite{Rei.85}. In addition to the short-range
correlations described explicitly by the unitary transformation, the RPA
ground state calculations allow for the inclusion of residual long-range 
correlations necessary to obtain a realistic, ab initio-type description of
nuclei.

On the other side, models based on phenomenological nuclear interactions
have been well established over the past decades, and have been very
accurately tuned to the properties of finite nuclei. These models are
nowadays successful not only in the region of stable nuclei, but also in the
description of exotic nuclear structure and collective excitations in
nuclei away from the valley of $\beta$-stability, both in the
non-relativistic \cite{Mat.01,Ter.04,Sar.04} and relativistic framework
\cite{Paa.03,Cao.05}. Particularly interesting is the phenomenon
of the pygmy dipole resonance in neutron-rich nuclei (PDR), indicating
that the loosely bound neutrons might coherently oscillate against the
approximately isospin-saturated proton-neutron core \cite{Vrepyg.01}. Very
recently it has been shown that for proton rich nuclei in the lower region
of the nuclide chart, one also could expect the appearance of a low-energy
exotic collective mode, i.e. the proton PDR, when loosely-bound protons
vibrate against the rest of the nucleons \cite{Paa_pp.05}. The experimental
evidence  about low-lying excitations in nuclei towards the drip-lines is
still rather limited, and actually available only in light nuclei up to
the oxygen isotopes \cite{Lei.01}. The present article will summarize the
recent progress in studies of low-lying excitations towards the
proton drip-line within the Relativistic Quasiparticle RPA (RQRPA).

\section{\label{secII}The Hartree-Fock model in the UCOM framework}

The UCOM approach aims at an explicit treatment of interaction-induced
short-range central and tensor correlations in nuclei
\cite{Fel.98,Nef.03,Rot.04,Rot.05}. These correlations are imprinted into an
uncorrelated many-body state $\ket{\Psi}$ through a state-independent
unitary transformation defined by the unitary operator $\CO$, resulting in
a correlated state,  $\ket{\corr{\Psi}} = \CO\; \ket{\Psi}.$ Due to the
unitarity of the correlation operator, matrix elements of an operator
$\OO$ in correlated many-body states are equal to those evaluated using
the correlated operator $\corr{\OO}$ and uncorrelated many-body states,
i.e.
\begin{equation}
\matrixe{\corr{\Psi}}{\OO}{\corr{\Psi}'}=
\matrixe{\Psi}{\CCO \OO \,\CO}{\Psi'}= 
\matrixe{\Psi}{\corr{\OO}}{\Psi'}.
\end{equation}
The short-range central and tensor correlations are separately included
via the unitary operators $\CO_{r}$ and $\CO_{\Omega}$, respectively, 
and formulated as exponential functions of the two-body Hermitian 
generators $\gO_r$ and $\gO_{\Omega}$. The operator form of the generators
is motivated by the basic physics of the two types of correlations we are
going to describe explicitly. 

The short-range central correlations are induced by the strong short-range
repulsion in the central part of realistic NN interactions. This repulsive
core prevents nucleons in a many-body system to approach each other closer
than the characteristic size of the core. In the two-body density matrix
these correlations are revealed through the depletion of the probability
density for particle distances smaller than the core radius. In order to
include these correlations into an uncorrelated many-body state, e.g., the
Slater determinant of the Hartree-Fock approach, we perform a
distance-dependent radial shift with respect to the relative coordinate of
two-particles. The corresponding Hermitian generator $\gO_r = \frac{1}{2}
[ s(\rO) \qO_r + \qO_r s(\rO) ]$ contains the radial component of the
relative momentum operator, $\qO_r$, and a function $s(r)$ which controls
the distance-dependence of the shift. Implementation of the correlation
operator in coordinate representation corresponds to a norm-conserving
coordinate transformation $\rV \mapsto \Rm(r) \frac{\rV}{r}$ of the
relative coordinate. The radial correlation function $\Rm(r)$ and its
inverse $\Rp(r)$ are related to the shift function $s(r)$
\cite{Rot.04,Rot.05}.

For the following calculations the Argonne V18 (AV18)
potential \cite{Wir.95} is used. In practice, the correlation functions
$\Rp(r)$ are parameterized and the optimal parameters are determined
for each spin-isospin channel from an energy minimization in the
two-body system \cite{Rot.05}.
Since the $(S=1, T=1)$-channel of the AV18 potential is
purely repulsive, we employ a simple constraint on the range of the
correlation function in this channel in order to avoid artificial
long-range correlation functions. We have checked explicitly that
the effect of variations of this constraint around the value that ensures
the short-range of the correlation function, is negligible for all
calculations presented here. 

A second, equally important type of correlations is induced by the tensor
part of the interaction \cite{Nef.03}. They entangle the relative spatial orientation of
two nucleons with their spin orientation. The generator $\gO_{\Omega}$
has to describe an angular shift depending on the spin orientation. 
This is achieved by  $\gO_{\Omega} =
\frac{3}{2}\vartheta(\rO)
[(\sigmaOV_1\!\cdot\qOV_{\Omega})(\sigmaOV_2\!\cdot\rOV)  +
(\sigmaOV_1\!\cdot\rOV)(\sigmaOV_2\!\cdot\qOV_{\Omega}) ]$, where
$\qOV_{\Omega} = \qOV - \tfrac{\rOV}{\rO}\qO_r$ \cite{Nef.03}.
As for the central correlations, the tensor correlation functions
$\vartheta(\rO)$, which control the distance-dependence of the tensor
correlator, are parameterized and determined by a two-body energy
minimization. A characteristic of the tensor part of realistic NN interactions
and  thus of tensor correlations is their long range. We are not aiming at a
description of long-range tensor correlations by the unitary
transformation, since they are strongly system-dependent. In contrast
to the deuteron, the long-range tensor correlations in heavier nuclei will
be largely screened. During the determination of the tensor correlators we
therefore constrain the range or volume of $\vartheta(r)$ given by
\eq{ \label{tensor_constraint}
  I_{\vartheta}^{(ST)} = \int\dd{r}\; r^2 \vartheta(r) .
}
In the present study, we use the optimal tensor correlators for
$I^{(S=1,T=0)}_{\vartheta}$ = 0.07, 0.08, and 0.09 fm$^3$ and investigate
the effect on the global properties of collective excitation phenomena. No
tensor correlator is employed in the ($S=1,T=1$)-channel, because there
the tensor interaction is rather weak. It has been shown within no-core
shell model calculations that for $I^{(10)}_{\vartheta}$ = 0.09 fm$^3$ 
the experimental binding energies for $A\leq4$ are reproduced
quite well \cite{Rot.05}.

The correlated operators constructed by the unitary transformation 
contain irreducible contributions, not only of one-body and two-body
operators, but also higher $n$-body parts. This cluster expansion is
truncated after the two-body level, leading to the so-called two-body
approximation. In previous studies it has been verified that higher order
contributions due to central correlations can be neglected in the
description of nuclear structure properties \cite{Rot.04}. For the tensor
correlators the range constraint is important for the validity of the
two-body approximation. The size of residual three-body and
higher order contributions was estimated in \cite{Rot.05}. 

Starting from the uncorrelated Hamiltonian for the $A$-body system,
consisting of the kinetic energy operator and the bare AV18 potential, 
the central and tensor correlation operators are employed to construct 
the correlated Hamiltonian in the two-body approximation. Therein the
one-body contributions come only from the uncorrelated kinetic energy,
while two-body contributions arise from the correlated kinetic energy and
the correlated potential, which together constitute the low-momentum
correlated interaction $\VO_{\UCOM}$ \cite{Rot.04,Rot.05}.  The
$\VO_{\UCOM}$ interaction can directly be employed in the HF model to
determine the single-particle wave functions and energies.  By using the
expansion in the harmonic oscillator basis, the HF equations are solved in
a self-consistent way,  with restrictions on the maximal value of the
major shell quantum number $N_{\max} = 12$, and maximal orbital angular
momentum quantum number $l_{\max}=8$. 

\begin{figure}
\includegraphics[width=0.9\textwidth]{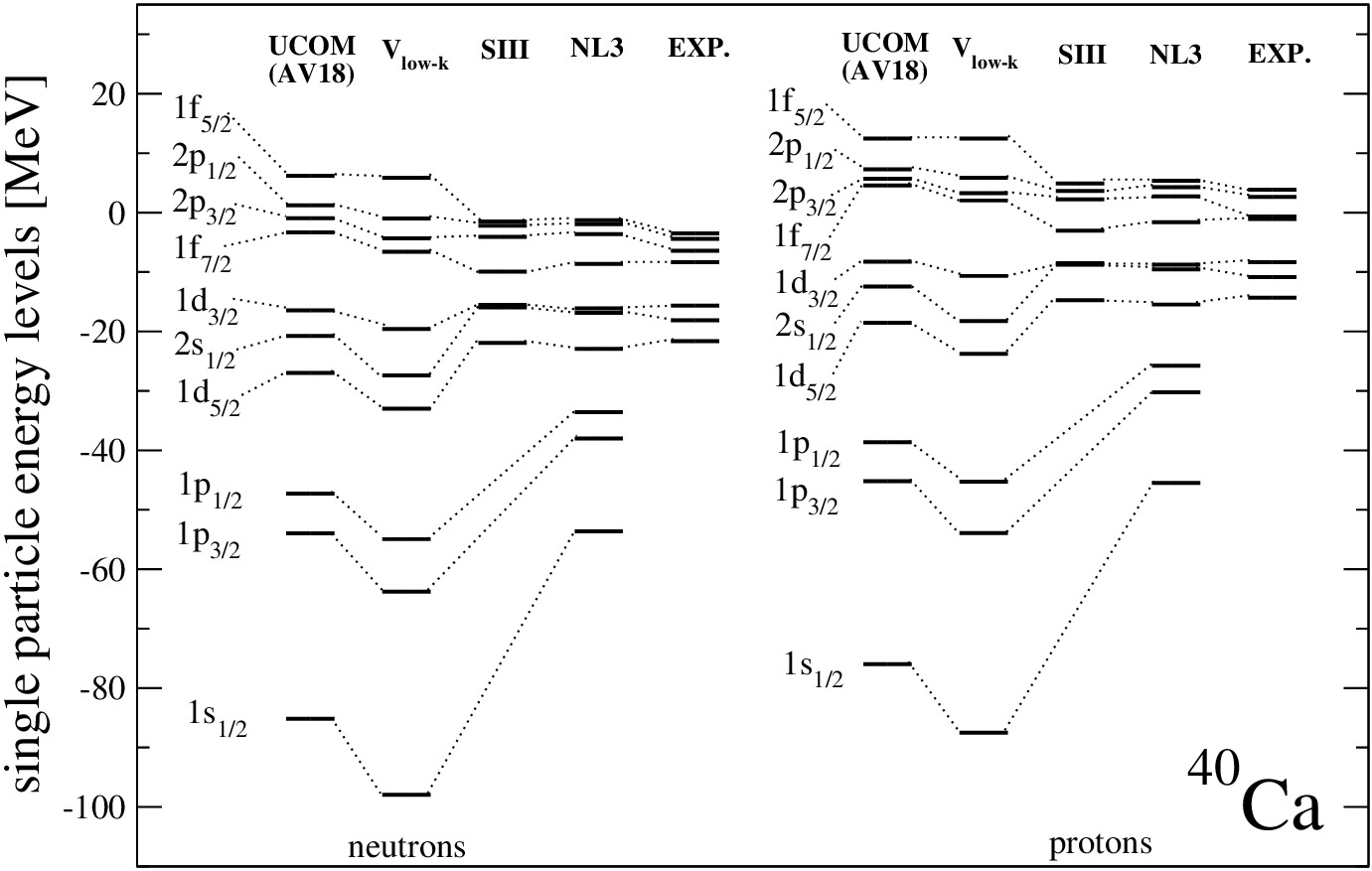}
\caption{The UCOM-HF neutron and proton single particle spectrum for $^{40}$Ca,
along with the corresponding spectra from the HF model based on
the low-momentum NN potential $V_{{\rm low}-k}$ \protect\cite{Cor.03}, 
HF with SIII Skyrme-type interaction \protect\cite{Isa.02}, relativistic mean field theory
with NL3 effective interaction \cite{LKR.97}, and experimental spectrum \cite{Isa.02}.
The UCOM-HF calculations are based on the correlated Argonne
V18 interaction ($I_{\vartheta}^{(10)}$=0.09 fm$^3$).
}
\label{figspectra2}
\end{figure}

In Fig. \ref{figspectra2} the UCOM-HF single-nucleon spectra are displayed
for the case of $^{40}$Ca. The calculations are based  on the correlated
Argonne V18 interaction, with the constraint $I_{\vartheta}^{(10)}$=0.09
fm$^3$ for the correlation volume of  the tensor correlator,
Eq.~(\ref{tensor_constraint}). The UCOM-HF energy levels are compared with
the HF  spectrum obtained with the low-momentum NN potential $V_{{\rm
low}-k}$ \cite{Cor.03}, with two standard phenomenological interactions in
the nonrelativistic (Skyrme) \cite{Isa.02} and relativistic (NL3)
\cite{LKR.97} framework, and with experimental levels \cite{Isa.02}. The
spectra obtained from the HF model based on realistic NN interactions
appear distributed too wide in  energy. In addition, the HF binding
energies and radii are too small compared to the experimental values
\cite{Rot.04}. 

These deviations can be attributed to several missing pieces in the
UCOM-HF description: (i) Long-range correlations are not covered by the
unitary correlation operators and should be described by the model space,
i.e. the available many-body states. The independent-particle states of
the HF approach are clearly not able to do so and one has to go beyond the
mean-field level. In the next section we are going to include long-range
correlations due to collective vibrations, by means of RPA.  (ii)
Three-body forces, either genuine or induced by the unitary
transformation, generally play a role for the quantitative description of
nuclear structure. These are not included in the present study. Hence the
results presented here will provide some information on their importance.

\section{Random-phase approximation based on the $\VO_{\UCOM}$}

The HF description of the nuclear ground state is to some extent
oversimplified,  and correlation effects going beyond mean-field
should be included.  Giant resonances may have some influence on the
nuclear binding energies \cite{Rei.85}, and it is known that correlations
due to surface vibrations have a considerable influence on the ground
state  densities \cite{Esb.83}.  In this section, we will employ an RPA
model based on the UCOM Hamiltonian (UCOM-RPA) to evaluate the ground
state correlations due to collective vibrations and to study the 
properties of such excitations themselves.

The UCOM-HF single-particle states  are used for the construction of the
$ph$ configuration space  for the RPA model. One of the standard
approaches to derive the RPA equations is the equation of motion method
with the quasiboson approximation \cite{Row.70}, resulting in the
eigenvalue problem formulated as a set of coupled equations for the 
forward and backward amplitudes, $X_{ph}^{k,JM}$ and $Y_{ph}^{k,JM}$
respectively,
\begin{equation}
\label{rpaeq}
\left(
\begin{array}{cc}
A^J & B^J \\
B^{^\ast J} & A^{^\ast J}
\end{array}
\right)
\left(
\begin{array}{c}
X^{k ,JM} \\
Y^{k,JM}
\end{array}
\right) =\omega_{k}\left( 
\begin{array}{cc}
1 & 0 \\
0 & -1
\end{array}
\right)
\left( 
\begin{array}{c}
X^{k,JM} \\
Y^{k,JM}
\end{array}
\right)\; .
\end{equation}
The eigenvalues $\omega_{k}$ correspond to RPA excitation energies and
the RPA matrices are given by,
\begin{eqnarray}
A^{J}_{php'h'} & = & \bra{\phi}\left[ \left[A_{ph}^{JM},\HO_{\UCOM}\right],{A_{p'h'}^{JM}}^{+} \right] \ket{\phi} \\
B^{J}_{php'h'} & = & -\bra{\phi}\left[ \left[A_{ph}^{JM},\HO_{\UCOM}\right],(-1)^{J-M}{A_{p'h'}^{J-M}} \right] \ket{\phi},
\end{eqnarray}
where the operator ${A_{ph}^{JM}}^{+}$ (${A_{ph}^{JM}}$) creates
(annihilates) a $ph$ state of angular momentum $JM$. We consistently use
the intrinsic Hamiltonian $\HO_{\UCOM}=\TO - \TO_{\cm} + \VO_{\UCOM}$,
i.e., the center of mass contribution to the kinetic energy is subtracted
on the operator level. The Coulomb interaction is included explicitly. 

An essential property of the present model is that it is fully
self-consistent,  i.e. the same correlated realistic NN interaction
$\VO_{\UCOM}$ is used in the HF equations that determine the
single-particle basis, and in the RPA residual interaction entering the
calculation of the   RPA matrices. This essential property of our model
ensures that RPA amplitudes do not contain spurious components associated
with the center-of-mass translational motion. We have verified that the
spurious $1^-$ state is properly decoupled from the physical excitation
states. We also have examined, for closed-shell
nuclei across the nuclide chart, that the UCOM-RPA model essentially
exhausts the  isoscalar energy-weighted sum rules \cite{Row.70} with
maximal discrepancies of $~\pm 3\%$. 

\begin{figure}
\includegraphics[width=0.9\textwidth]{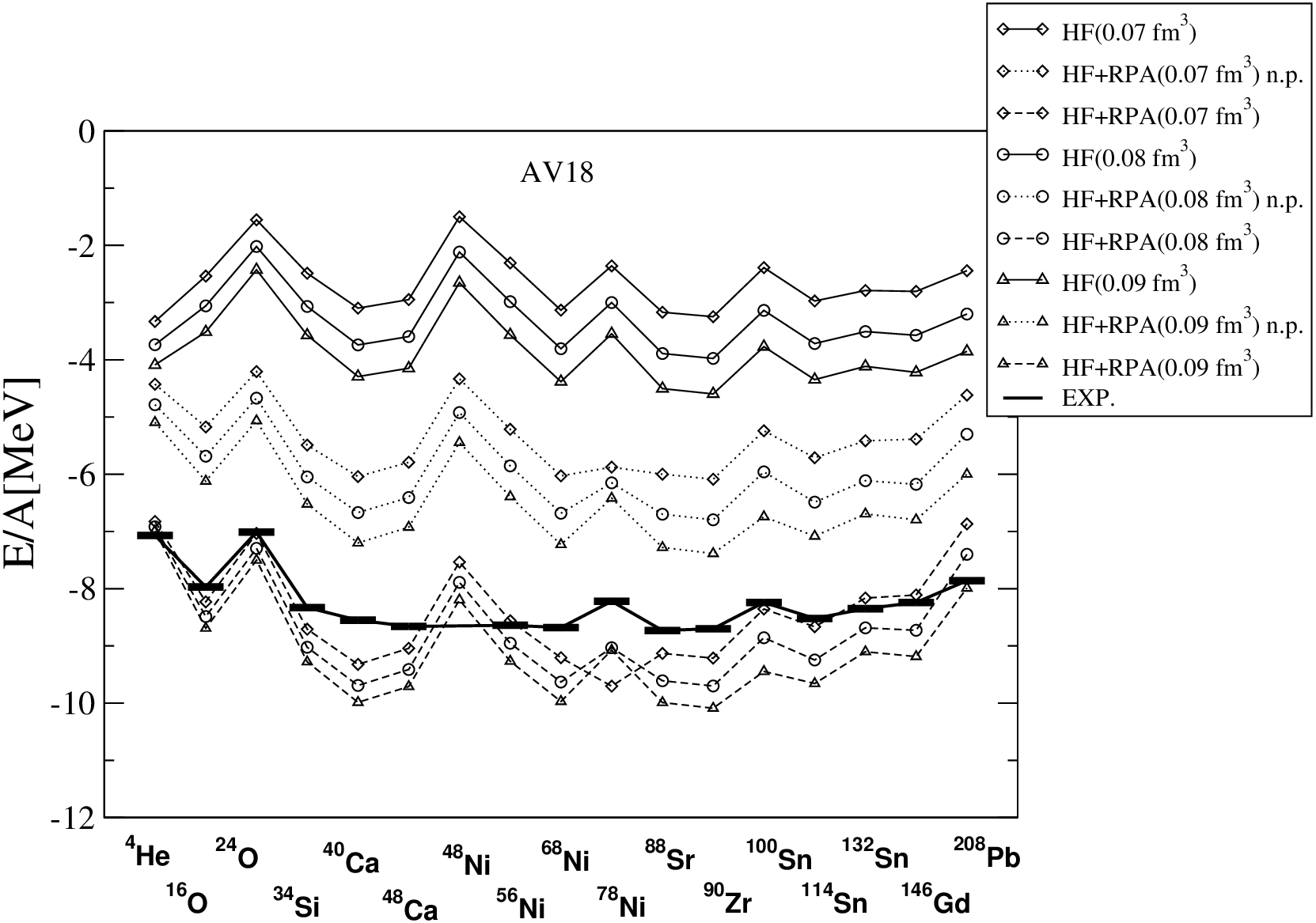}
\caption{The Hartree-Fock binding energies for a series of closed-shell nuclei, in comparison with the 
corrected binding energies due to UCOM-RPA correlations from collective vibrations, and experimental 
data \protect\cite{Aud.95}. The binding energies with the full correction (dashed line) and only with natural
parity excitations (dotted line, n.p.) are separately displayed. Both the Hartree-Fock and RPA calculations are
based on the correlated Argonne V18 interaction 
with various ranges of the tensor correlator ( $I_{\vartheta}^{(10)}$=0.07,
0.08, and 0.09 fm$^3$). }
\label{figcorren2}
\end{figure}

In the present study,
the correlation energies are evaluated within the UCOM-RPA framework,
\begin{equation}
\delta E = -\sum_{k,J} (2J+1) \hbar {\omega}^{J}_{k}\sum_{ph}|Y^{k,J}_{ph}|^2,
\label{corren1}
\end{equation}
by using the RPA eigenvalues $\omega_{k}^{J}$, and backward-going
amplitudes $Y^{k,J}_{ph}$. Both, the natural $\pi=(-1)^J$ and  unnatural
parity $\pi=(-1)^{J+1}$ excitations are included, in the range of
$J^{\pi}=0^{\pm}-10^{\pm}$. The HF binding energies together with RPA
correlations due to collective excitations are shown in
Fig.~\ref{figcorren2} for several closed-shell nuclei. The HF binding
energies and UCOM-RPA correlations are calculated in a consistent way by
using the correlated AV18 interaction for various ranges of the tensor
correlator, constrained by $I_{\vartheta}^{(10)}$=0.07, 0.08, and
0.09 fm$^3$.  In general, the tensor correlator with longer range provides
stronger binding both on the HF level and when the correlations are taken
into account. In comparison with the experimental binding energies
\cite{Aud.95}, the present model with full implementation of RPA
correlations  seems to favor the tensor correlator with shorter range. 
However, one should keep in mind that the method used to evaluate the
correlation energy, Eq.~(\ref{corren1}), is not free of over-counting \cite{Ula.69}
and therefore the correlation effects are overestimated. Within 
many-body perturbation theory or configuration interaction calculations,
the longer ranged correlator provides a very good agreement with
experimental binding energies for all nuclei in accord with the no-core
shell model calculations discussed in \cite{Rot.05}.

\begin{figure}
\includegraphics[width=0.9\textwidth]{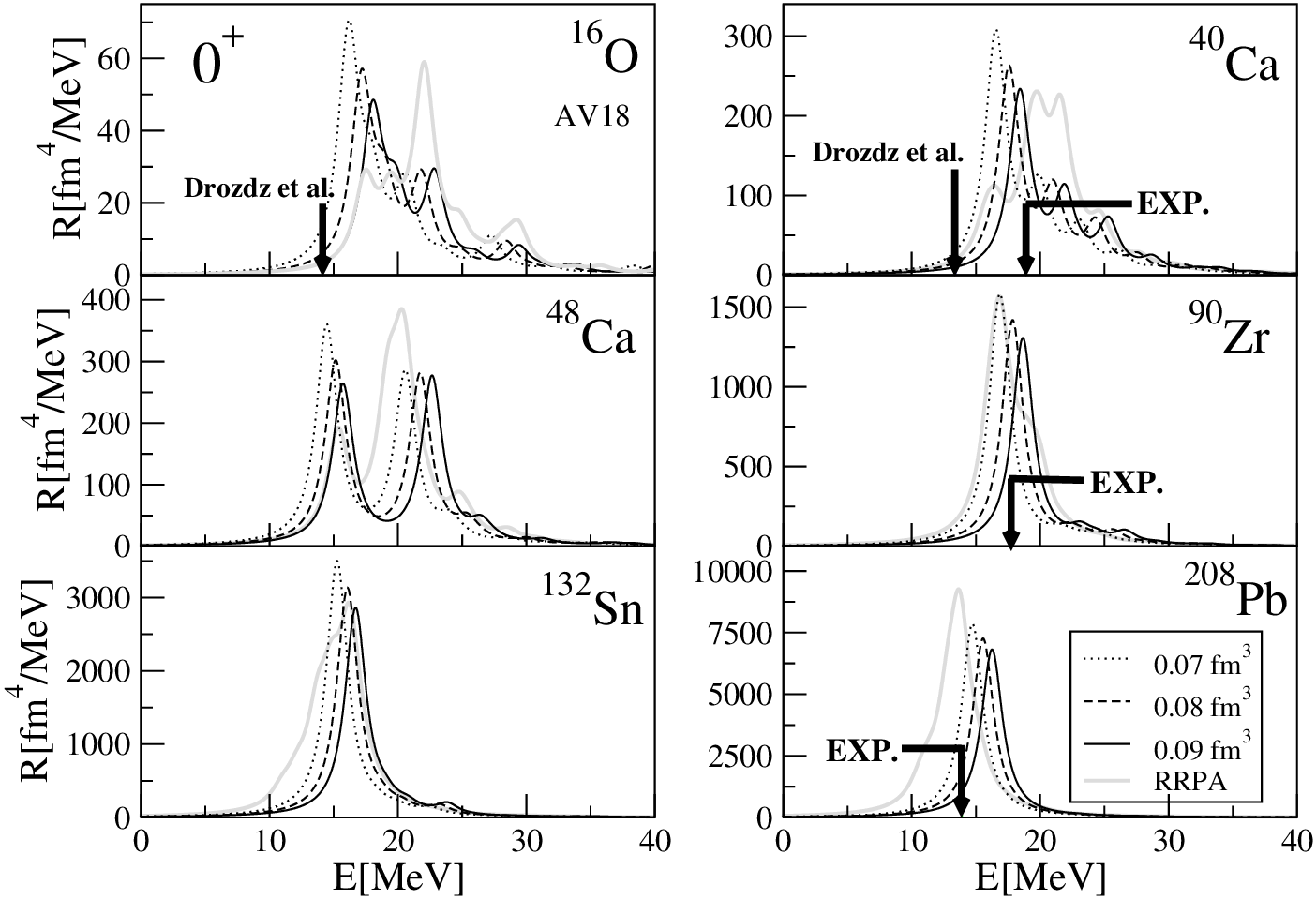}
\vspace{1cm}
\caption{The UCOM-RPA monopole transition strength distributions for the
correlated Argonne V18 interaction, using different restrictions 
on the range of the tensor correlator ($I_{\vartheta}^{(10)}$=0.07, 0.08,
and 0.09 fm$^3$). The grey lines correspond to the monopole response
from the relativistic RPA, based on the effective Lagrangian with
density-dependent meson-nucleon couplings \protect\cite{Nik1.02},
with DD-ME1 parameterization \protect\cite{Nik2.02}.
The ISGMR centroid energies obtained from 
nonrelativistic (Dro{\. z}d{\. z} et al.) calculations \protect\cite{Dro.90} 
and experimental data \protect\cite{You.99,Shl.93,Sha.88} 
are denoted by arrows.}
\label{figmono2}
\end{figure}

By examining the electric transition strength, one can also study the
properties of the correlated interaction $\VO_{\UCOM}$.  This is
exemplified in Fig.~\ref{figmono2}, where the UCOM-RPA strength
distributions, corresponding to the isoscalar giant monopole resonance
(ISGMR) are displayed for the correlated AV18 interaction with different
restrictions on the range of the tensor correlator,
$I_{\vartheta}^{(10)}$=0.07, 0.08, and 0.09 fm$^3$. For the lighter nuclei
$^{16}$O, $^{40}$Ca, and $^{48}$Ca, the ISGMR is fragmented into two-three
peaks, whereas for $^{90}$Zr, $^{132}$Sn, and $^{208}$Pb the  ISGMR is
strongly collective, resulting essentially in a single peak. For a comparison,
the monopole response is also calculated in the framework of relativistic RPA
based on effective Lagrangian with density-dependent meson-nucleon
vertex functions, with DD-ME1 interaction (more details are given in
Sec. \ref{secIV}). In addition, the calculated ISGMR strength distributions
are compared with the nonrelativistic RPA based on
Woods-Saxon potential and G-matrix formalism \cite{Dro.90}, and 
with experimental data from $(\alpha,\alpha)$ \cite{You.99,Shl.93} and
($^3$He,$^3$He)  scattering \cite{Sha.88}.
One can observe that a decrease of the range of the tensor correlator
systematically pushes the transition
strength towards lower energies. In particular, by decreasing the range of
the tensor correlator, i.e. its constraint $I_{\vartheta}^{(10)}$=0.09 fm$^3$ towards
0.07 fm$^3$, the excitation energy of ISGMR lowers by $\approx$4 MeV. This
means that by varying the range of the tensor  correlator, one can
effectively control the impact of the missing correlations on the
transition strength of ISGMR.

\section{\label{secIV}Proton pygmy dipole resonance in the relativistic QRPA}

In this section we discuss recent  developments regarding the response of
nuclei far from $\beta$-stability.  One of the major challenges in this
region  is the understanding of soft modes of excitations which involve
loosely bound nucleons.  In particular, in neutron-rich nuclei, nucleons
from the neutron skin may give rise to a soft low-energy dipole mode known
as the  pygmy dipole resonance (PDR)~\cite{Suz.90,Paar.05,Paar_02.05}. The
structure of nuclei on the proton-rich side is equally important for many
aspects of the underlying many-body problem and effective nuclear
interactions. In a recent relativistic QRPA study, it has been predicted
that in nuclei with proton excess one could expect the appearance of
a proton PDR mode where loosely bound protons vibrate against the rest of
the nucleons \cite{Paa_pp.05,Paa_pp2.05}.  The relativistic QRPA \cite{Paa.03}
is formulated in the canonical single-nucleon basis of the relativistic
Hartree-Bogoliubov (RHB) model and is fully self-consistent. For the
interaction in the particle-hole channel effective Lagrangians with
nonlinear meson self-interactions or density-dependent meson-nucleon
couplings are used \cite{Nik1.02}, and pairing correlations are described
by the pairing part of the finite-range Gogny interaction \cite{Dec.80}.
The parameters of the effective relativistic interaction have been
adjusted to properties of nuclear matter and to binding energies, charge
radii, and differences between neutron and proton radii of spherical
nuclei~\cite{Nik1.02}. In the small-amplitude limit, the RQRPA equations
are derived from the equation of motion for the generalized nucleon
density \cite{Paa.03}. The RQRPA configuration space is constructed from
standard ($2qp$) pairs, but one also needs to include transitions to the
unoccupied states from the Dirac sea \cite{Paar.04,Vre.05}. 

\begin{figure}
\includegraphics[width=0.9\textwidth]{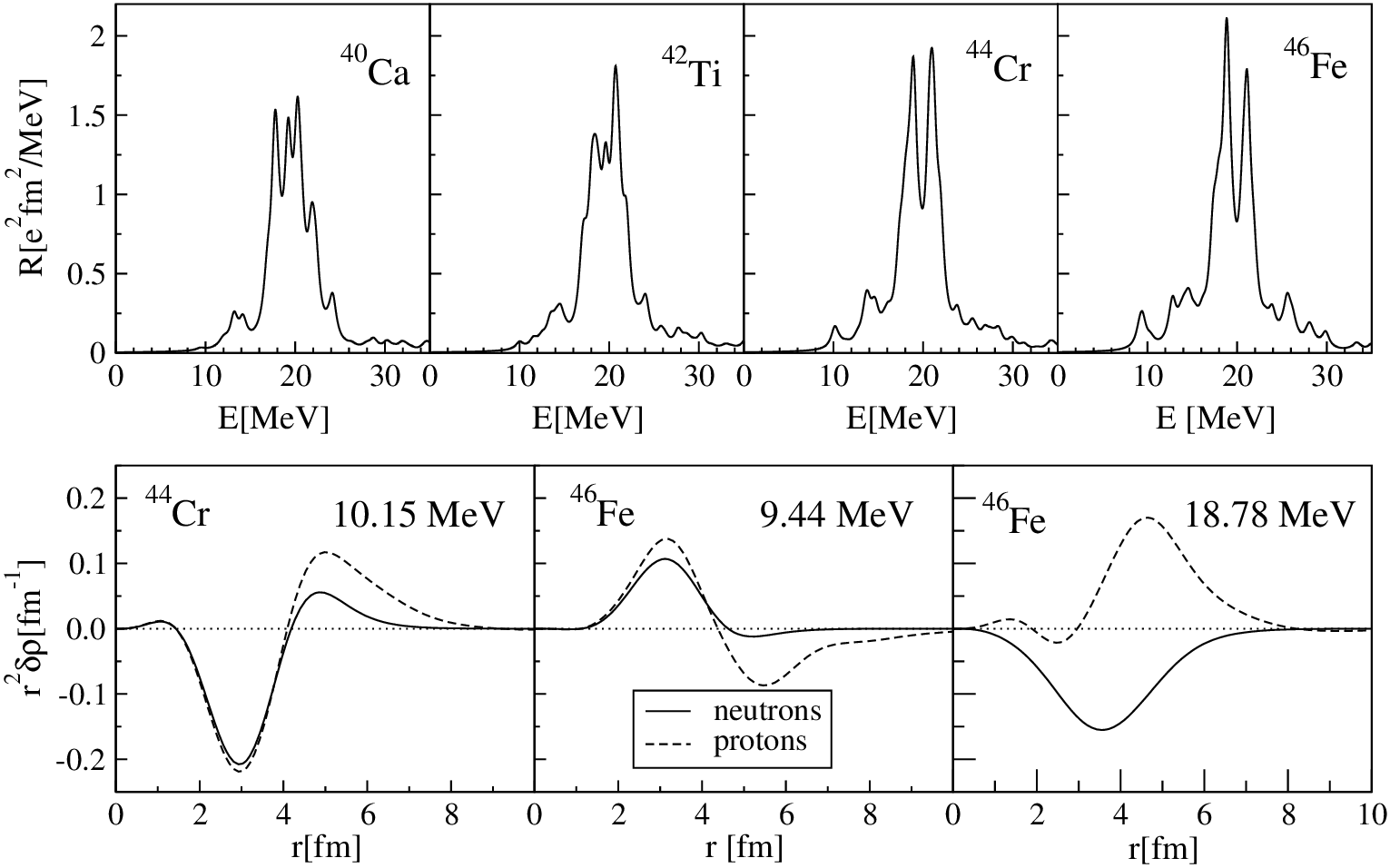}
\caption{The RHB+RQRPA isovector dipole strength distributions 
in the N=20 isotones, calculated with the 
DD-ME1 effective interaction. For $^{44}$Cr and $^{46}$Fe the
proton and neutron transition densities for the main
peak in the low-energy region are displayed in the 
lower panel and, for $^{46}$Fe, the transition densities 
for the main GDR peak.}
\label{fig1}
\end{figure}

The RQRPA dipole strength distributions for N=20 isotones, displayed in
Fig.~\ref{fig1}, are dominated by the  isovector giant dipole resonances
(GDR) at $\approx 20$ MeV excitation energy. With the increase of the
number of protons,   low-lying dipole strength appears in the region below
the GDR and,  for $^{44}$Cr and $^{46}$Fe, a pronounced  low-energy peak
is found at $\approx 10$ MeV excitation energy.  In the lower panel of
Fig.~\ref{fig1} we plot the proton and neutron  transition densities for
the peaks at 10.15 MeV in $^{44}$Cr and  9.44 MeV in $^{46}$Fe, and
compare them with the transition densities of the GDR state at 18.78 MeV
in $^{46}$Fe. Obviously the dynamics  of the two low-energy peaks is very
different from that of the isovector GDR: the proton and neutron
transition densities are in phase in the nuclear interior and  there is
very small contribution from the neutrons in the surface region. By
exploring the RQRPA amplitudes, we note that, rather than a single proton
{$2qp$} excitation, the low-lying states are characterized  by a
superposition of a number of mainly proton {$2qp$} configurations. The
low-lying state does not belong to statistical E1 excitations sitting on
the tail of the GDR, but represents a fundamental mode of excitation: the
proton electric pygmy  dipole resonance (PDR).

\section{\label{secVI}Summary}

In the present study, a fully self-consistent RPA model is constructed in
the single-nucleon Hartree-Fock basis, by using correlated  realistic NN
interactions obtained within the UCOM framework. It is shown that the
$V_{\rm UCOM}$ interaction generates a strongly collective ISGMR mode, 
whose energy is sensitive to the range of the tensor correlator. The
UCOM-RPA correlations due to collective vibrations provide important
contributions to the nuclear binding energies. In addition, by employing
the fully  self-consistent relativistic quasiparticle RPA, it is indicated
that the nuclei towards the proton drip-line are characterized by the
appearance of the proton pygmy dipole resonance, i.e. an exotic mode where
loosely bound protons oscillate against the isospin-saturated
proton-neutron core.

\leftline{\bf ACKNOWLEDGMENTS}
This work has been supported by
the Deutsche Forschungsgemeinschaft (DFG) under contract SFB 634.


%
\newpage

\bigskip
\begin{thebibliography}{999}

\bibitem{Vau.72} D. Vautherin and D. M. Brink, Phys. Rev. C {\bf 5}, 626 (1972).
\bibitem{Dec.80} J. Decharg{\'e} and D. Gogny, Phys. Rev. C {\bf 21}, 1568 (1980).
\bibitem{Vre.05} D. Vretenar, A. V. Afanasjev, G. A. Lalazissis, and P. Ring,
Phys. Rep. {\bf 409}, 101 (2005).
\bibitem{Mac.01} R. Machleidt, Phys. Rev. C {\bf63}, 024001 (2001).
\bibitem{Wir.95} R. B. Wiringa, V. Stoks, and R. Schiavilla, Phys. Rev. C {\bf 51}, 38 (1995).
\bibitem{Ent.02} D. R. Entem and R. Machleidt, Phys. Lett. B {\bf 524}, 93 (2002).
\bibitem{Pie.04} S. C. Pieper, R. B. Wiringa, and J. Carlson, Phys. Rev. C {\bf 70}, 054325 (2004).
\bibitem{Nav.00} P. Navr{\'a}til, J. P. Vary, and B. R. Barrett, Phys. Rev. C {\bf 62}, 054311 (2000).
\bibitem{Pie.01} Steven C. Pieper, et al., Phys. Rev. C {\bf 64}, 014001 (2001).
\bibitem{Epe.02} E. Epelbaum et al., Phys. Rev. C {\bf 66}, 064001 (2002).
\bibitem{Fel.98} H. Feldmeier, T. Neff, R. Roth, J. Schnack, Nucl. Phys. A
{\bf 632}, 61 (1998). 
\bibitem{Nef.03} T. Neff and H. Feldmeier, Nucl. Phys. A {\bf 713}, 311(2003).  
\bibitem{Rot.04} R. Roth, T. Neff, H. Hergert, and H. Feldmeier, Nucl. Phys. A {\bf 745}, 3 (2004).
\bibitem{Rot.05} R. Roth, H. Hergert, P. Papakonstantinou, T. Neff, and H. Feldmeier, nucl-th/0505080 (2005).
\bibitem{Bog.03} S.~K. Bogner, T.~T.~S. Kuo, A. Schwenk, Phys. Rep. {\bf 386}, 1 (2003).
\bibitem{Row.70} D. J. Rowe, Nuclear Collective Motion, (Methuen and Co. LTD., London 1970).
\bibitem{Rei.85} P.-G. Reinhard and J. Friedrich, Z. Phys. A {\bf 321}, 619 (1985).
\bibitem{Mat.01} M. Matsuo, Nucl. Phys. A {\bf 696}, 371 (2001).
\bibitem{Ter.04} J. Terasaki, J. Engel, M. Bender, J. Dobaczewski, W. Nazarewicz, and
M. Stoitsov, Phys. Rev. C {\bf 71}, 034310 (2005).  
\bibitem{Sar.04} D. Sarchi, P. F. Bortignon, and G. Col\'o,
        Phys. Lett. B {\bf 601}, 27 (2004).
\bibitem{Paa.03} N. Paar, P. Ring, T. Nik\v si\' c, and D. Vretenar,
        Phys. Rev. C {\bf 67}, 034312 (2003).
\bibitem{Cao.05}L. G. Cao and Z. Y. Ma, Phys. Rev. C {\bf 71},  034305 (2005). 
\bibitem{Vrepyg.01} D. Vretenar, N. Paar, P. Ring, and G. A. Lalazissis
Nucl. Phys. A {\bf 692}, 496 (2001).
\bibitem{Paa_pp.05} N. Paar, D. Vretenar, and P. Ring, Phys. Rev. Lett. {\bf 94}, 182501 (2005).
\bibitem{Lei.01} A. Leistenschneider et al., Phys. Rev. Lett. {\bf 86}, 5442 (2001).
\bibitem{Cor.03} L. Coraggio et al., Phys. Rev. C {\bf 68}, 034320 (2003).
\bibitem{Isa.02} V. I. Isakov et al., Eur. Phys. J. A {\bf 14}, 29 (2002).
\bibitem{LKR.97} G.A. Lalazissis, J. K\"onig, and P. Ring,
Phys. Rev. C {\bf 55}, 540 (1997).
\bibitem{You.99} D. H. Youngblood, H. L. Clark, and Y.-W. Lui, 
Phys. Rev. Lett. {\bf 82}, 691 (1999).
\bibitem{Shl.93} S. Shlomo and D. H. Youngblood, Phys. Rev. C {\bf 47}, 529 (1993).
\bibitem{Sha.88} M. M. Sharma and M. N. Harakeh, Phys. Rev. C {\bf 38}, 2562 (1988).
\bibitem{Dro.90} S. Dro{\. z}d{\. z}, S. Nishizaki, J. Speth, and J. Wambach,
Phys. Rep. {\bf 197}, 1 (1990).
\bibitem{Esb.83} H. Esbensen and G. F. Bertsch, Phys. Rev. C {\bf 28}, 355 (1983).
\bibitem{Aud.95} G. Audi and A. Wapstra, Nucl. Phys. A {\bf 595}, 409 (1995).
\bibitem{Ula.69} N. Ullah and D. J. Rowe, Phys. Rev. {\bf 188}, 1640 (1969).
\bibitem{Suz.90} Y. Suzuki, K. Ikeda, and H. Sato,
        Prog. Theor. Phys. {\bf 83}, 180 (1990).
\bibitem{Paar.05} N. Paar, T. Nik\v si\' c, D. Vretenar, and P. Ring,
        Phys. Lett. B {\bf 606}, 288 (2005).
\bibitem{Paar_02.05} N. Paar,  T. Nik\v si\' c, D. Vretenar, and P. Ring,
Int. J. Mod. Phys. E {\bf 14}, 1 (2005).
\bibitem{Paa_pp2.05} N. Paar, P. Papakonstantinou,
V. Yu. Ponomarev, and J. Wambach, nucl-th/0506010 (2005).
\bibitem{Nik1.02} T. Nik\v si\' c, D. Vretenar, P. Finelli, and P. Ring,
        Phys. Rev. C {\bf 66}, 024306 (2002). 
\bibitem{Nik2.02} T. Nik\v si\' c, D. Vretenar, and P. Ring,
        Phys. Rev. C {\bf 66}, 064302 (2002). 
\bibitem{Paar.04} N. Paar,  T. Nik\v si\' c, D. Vretenar, and P. Ring, 
Phys. Rev. C {\bf 69}, 054303 (2004).
\end{thebibliography}
\end{document}